\documentstyle[12pt,epsf]{article}

        \def\l{\label}
         \def\d{\dagger}
         \def\be{\begin{equation}}
         \def\bea{\begin{eqnarray}}
          \def\la{\lambda}
         \def\o{\over}
         \def\b{\beta}
         \def\ro{\rho}
         \def\a{\alpha}
         \def\ee{\end{equation}}
         \def\eea{\end{eqnarray}}
         \def\R{\rm {I\kern-.200em R}}
         \def\C{\rm {I\kern-.520em C}}
         \def\r{\ref}
         \def\c{\cite}
\begin{document}
\begin{titlepage}
\vspace*{5mm}
\begin{center} {\Large \bf More on generalized simplicial chiral models }\\
\vskip 1cm
{\bf M. Alimohammadi}$^{a,b}$ \footnote {e--mail:alimohmd@theory.ipm.ac.ir},
{\bf Kh. Saaidi}$^a$ \footnote {e--maill:lkhled@molavi.ut.ac.ir}\\

\vskip 1cm
{\it $^a$Physics Department, University of Tehran, North Karegar Ave.,} \\
{\it Tehran, Iran }\\
{\it  $^b$Institute for Studies in Theoretical Physics and Mathematics,}\\
{\it  P.O.Box 5531, Tehran 19395, Iran}
\end{center}
\vskip 2cm
\begin{abstract}
By generalizing the auxiliary field  term in the Lagrangian of  simplicial chiral
models on a $(d-1)$--dimensional simplex, the generalized simplicial chiral
models has been introduced in \c{Ali}.
These models can be solved analytically only in $d=0$ and $d=2$
cases at large--N limit. In $d=0$ case, we calculate the eigenvalue
density function in strong regime and show that the partition
function computed from this density function is consistent with
one calculated by path integration directly. In $d=2$ case, it is
shown that all $V= {\rm Tr}(AA^{\d})^n$ models have a third order phase
transition, same as the 2--dimensional Yang--Mills theory.
\end{abstract}

\end{titlepage}
\newpage

\section{ Introduction }

The 1/N expansion of matrix--valued field theories is probably
the most important nonperturbative and nonnumerical theoretical
tool presently available in the study of such models as
non--Abelian gauge theories and two--dimensional quantum gravity.
The first major result concerning the large--N limit of matrix
theories was due to 't Hooft, who made the crucial observation
that in the 1/N expansion of continuum gauge theories, the set
of Feynman diagrams contributing to any given order admits a simple
topological interpretation \c{gt}.

Unfortunately, our knowledge of exact solution for the large--N
limit is limited to a small number of few--matrix systems, and it
becomes even smaller if we restrict ourselves to the case of
unitary matrix fields, which is relevant to the problem of lattice
QCD. After the solution of Gross and Witten \c{dg} of the single
link problem and its generalizations, exact results were obtained
only for the external field problem \c{rc1,eb} and a few toy models
(L=3,4 chiral chains)\c{rc2,df}. Therefore extending the number of
solved few--matrix systems is an important progress from different
points of view: first from pure theoretical reason. Second by noticing that
any few--matrix system have a reinterpretation, via double scaling limit, as some
different kind of matter coupled to two--dimensional quantum
gravity. And third because every few--matrix system involving unitary
matrices can be reinterpreted  as the generating functional for a
class of integrals over unitary groups, and these integrals in turn
are the essential missing ingredient in the context of a complete
algorithmization of the strong coupling expansion of many
interesting models \c{Cam}.

One of the interesting classes of the
few--matrix models is the simplicial chiral model (SCM)\c{pr2}. In this
model, each unitary matrix interacts in a fully symmetric way with
all other matrices, by preserving the global chiral invariance. The
resulting system can be described as a chiral model on $(d-1)$--dimensional
simplex. A simplex is formed by connecting $d$ vertices
by $d(d-1)/2$ links. In ref.\c{rc3}, where the large--N saddle--point
equations for density function $\rho(z)$ have been found, the
authors have introduced a single auxiliary variable $A$ (a complex
matrix) to decouple the matrix interaction, and by performing the
single--link external field integral, the saddle--point equation in
strong and weak regions have been found. In $d=2$, $d=4$, and $d
\rightarrow \infty$, the saddle--point equation has been solved
analytically and the phase transition of the model has been
specified. In the action of SCM in terms of the matrix $A$, this
matrix appears as a ${\rm Tr}(AA^{\d})$ term.

On the other hand, it is
well known that the pure 2--dimensional Yang--Mills theory (YM$_2$)
can be represented by the Lagrangian $i{\rm Tr}(BF)+{\rm Tr}(B^2)$, in which
$F$ is the usual field--strength tensor and $B$ is an auxiliary
pseudo--scalar field. Many of the main properties of YM$_2$ does
not change if one considers the generalized two--dimensional
YM$_2$ (gYM$_2$), which can be found by replacing ${\rm Tr}(B^2)$ term in
YM$_2$ action by an arbitrary class function $f(B)$\c{ew}. The
partition function of these theories have been fully studied in
different contexts in refs.\c{ew}--\c{mk} and the phase structure of
some of the specific examples has been studied in refs. \c{ma1} and \c{ma2}. In
all the studied cases, it is seen that the models have third--order
phase transition, which is the same behavior as YM$_2$. The crucial
point in this area is that such investigation for continuum gYM$_2$
is very complicated and there is no general result for the phase
transition of an arbitrary gYM$_2$ theory.

In ref. \c{Ali}, the procedure used to obtain the gYM$_2$ from
YM$_2$, i.e. ${\rm Tr}(B^2) \rightarrow f(B)$, has been used in SCM
to introduce the generalized simplicial chiral model (gSCM). That
is, the ${\rm Tr}(AA^{\d})$ term in SCM has been replaced by an arbitrary
class function $V(AA^{\d})$. The large--N limit of the model has
been discussed and the saddle--point equations in strong and weak regimes
have been found. Note that as the SCM at $d=2$ is the discrete version of
YM$_2$ \c{rc3}, we can consider the gSCM, at $d=2$, as the equivalent matrix
theory
of gYM$_2$. The phase structure of the model for $d=2$ and $V= {\rm Tr}(AA^{\d})^n$
($n=2,3,4$) is also obtained in \c{Ali}.

In this paper we want to complete our investigation of $d=2$ gSCM
and show that all $V={\rm Tr}(AA^{\d})^n$ models (with $n$ an arbitrary
positive integer) have third--order phase transition. We think that
it is an important result as it indirectly proves the equivalence of
all ${\rm Tr}(B^n)$ gYM$_2$ theories with YM$_2$, from the phase
structure point of view. The plane of the paper is as follows. In
section 2, we bring a brief review of SCM and gSCM and also the
saddle--point equations in weak and strong regimes. In section 3, we
focus on $d=0$ case and show that the calculation of the partition
function of the model, described in terms of $A$ fields, leads to
the trivial result which obtained from the main action. This is a
consistency check of our procedure. Finally in section 4
we investigate the phase structure of $V={\rm Tr}(AA^{\d})^n$
gSCM in $d=2$ and show that all of these models have third--order
phase transition. We also discuss the variation of the
discontinuity of $\b^2\partial C(\b,N)/\partial \b$ with $n$
at the critical point ($C$ is the specific heat and $\b = (g^2N)^{-1}$,
where $g$ is the coupling constant).

\section{The gSCM }

If we assign a U($N$) matrix to each vertex of a $(d-1)$--dimensional simplex,
then the partition function of simplicial chiral models is defined by \c{pr2}
\be \l{1}
Z_d(\b ,N)=\int \prod_{i=1}^ddU_i{\rm exp} \{ N\b \sum_{i=1}^d\sum_{j=i+1}^d
{\rm Tr} (U_iU_j^{\d}+U_i^{\d}U_j)\}.
\ee
To find this partition function, the authors of \c{pr2} have introduced  a
single auxiliary variable $A$ to decouple the matrix
interactions. The resulting partition function is
\be \l{2}
Z_d=\tilde Z_d/\tilde Z_0,
\ee
where
\be \l{3}
\tilde Z_d=\int \prod_{i=1}^ddU_idA{\rm exp} \{-N\b{\rm Tr}AA^{\d}+N\b{\rm Tr}
A\sum_iU_i^{\d}+N\b{\rm Tr}A^{\d}\sum_iU_i-N^2\b d\}.
\ee
Performing the single--link integral over the matrices $U_i$
\be \l{4}
e^{NW(BB^{\d})}\equiv \int dU{\rm exp}[N{\rm Tr}(B^{\d}U+U^{\d}B)],
\ee
we obtain
\be \l{5}
\tilde Z_d=\int dA{\rm exp} \{-N\b{\rm Tr}AA^{\d}+NdW(\b^2AA^{\d})-N^2\b d\}.
\ee
The gSCM is defined through the action \c{Ali}
\be \l{6}
Z_{d,V}=\tilde Z_{d,V}/\tilde Z_{0,V} \;\; ,
\ee
in which
\be \l{7}
\tilde Z_{d,V}=\int dA{\rm exp} \{-N\b V(AA^{\d})+NdW(\b^2AA^{\d})-N^2\b d\},
\ee
where $V(AA^{\d})$ is an arbitrary class function of $AA^{\d}$:
\be \l{8}
V(AA^{\d})=\sum_{n=1}a_n{\rm Tr}(AA^{\d})^n.
\ee
In special case $\a_n = \delta_{n,1}$, the gSCM reduces to SCM.
If we denote the  eigenvalues of the Hermitian semi--positive--definite matrix
$\b^2AA^{\d}$ by $x_i$'s, and at the large--N limit, which we are
interested in, using the expression $W(x_i)$ in the strong
and weak regimes \c{dg,rc1}, it can be shown that the saddle--point equation
for the eigenvalue density function $\rho(z)$ ($z_i=\sqrt{x_i+c}$)
is \c{Ali}:
\be\l{9}
z\sum_{k=1}{{ka_k}\o \b^{2k-1}}(z^2-c)^{k-1}-d={1\o 2}{\cal P}\int^b_adz'\ro (z')
\left( {2\o {z-z'}}-{{d-2}\o {z+z'}}\right).
\ee
In the above equation ${\cal P}$ indicates the
principal value of integral, and   the parameters $a$
and $b$ must be determined dynamically. The normalization condition of $\ro (z)$
is
\be\l{10}
\int_a^b\ro (z')dz'=1.
\ee
In the weak coupling regime ($\b > \b_c$), $c_w=0$ and the density
function $\ro (z)$ must satisfy
\be\l{11}
\int_a^bdz'{{\ro (z')}\o {z'}}\leq 2.
\ee
In the strong coupling regime $(\b < \b_c)$, $c_s=a^2$  and $\ro (z)$
must satisfy
\be\l{12}
\int_a^bdz'{{\ro (z')}\o {z'}}= 2.
\ee
Using the standard method of solving the integral equation \c{ga},
it can be shown that the density function $\ro (z)$ in weak regime
satisfies the following equation
\bea\l{13}
\ro_w(z)={{\sqrt{(b-z)(z-a)}}\o \pi}&{\Biggr \{}&\sum_{m,p,q}{{ma_m}\o {\b^{2m-1}}}
C_pC_{2m-p-q-2}z^qa^pb^{2m-p-q-2} \cr
&&-{{d-2}\o 2}\int_a^b{{dy}\o {y+z}}{{\ro_w(y)}  \o {\sqrt{(b+y)(y+a)}}}{\Biggl
\}}, \ \ \  \ \ \ {\rm for} \ \ \b>\b_c,
\eea
where $C_n=(2n-1)!!/(2^nn!)$, and in strong regime satisfies
\bea\l{14}
\ro_s(z)=-{z\o \pi}\sqrt{{{b-z}\o {z-a}}}&{\Biggr \{}& \sum_{n,m,p,q}
{{na_n}\o {\b^{2n-1}}}(-a^2)^p {{n-}1 \choose p}B_mC_{2n-2p-m-q-2}z^qa^m
b^{2n-2p-m-q-2}  \cr \cr
&&+ {{d-2}\o 2}\int_a^b{{dy}\o {y+z}}
\sqrt{{{y+a}\o {y+b}}}{{\ro_s(y)}\o y} {\Biggl \}}, \ \ \  \ \ \ {\rm for} \ \ \b<\b_c,
\eea
where $B_m=(2m-3)!!/(2^mm!)$ ($B_0\equiv -1$) \c{Ali}. Also by
investigating the behavior of the integrals at $z \rightarrow
\infty$, it can be shown that the parameters $a$ and $b$ in $\b > \b_c$
must be determined from the following equations
\be\l{15}
\sum_{n,m}{{na_n}\o {\b^{2n-1}}}C_mC_{2n-m-1}a^mb^{2n-m-1}-2=0,
\ee
and
\be\l{16}
\sum_{n,m}{{na_n}\o {\b^{2n-1}}}C_mC_{2n-m}a^mb^{2n-m}-(a+b)=1,
\ee
and in  strong regime, $\b < \b_c$, from the equations
\be\l{17}
\sum_{n,m,p}{{na_n}\o {\b^{2n-1}}}(-a^2)^p{{n-1} \choose p}B_mC_{2n-2p-m-1}a^m
b^{2n-2p-m-1}+2=0,
\ee
and
\be\l{18}
-\sum_{n,m,p}{{na_n}\o {\b^{2n-1}}}(-a^2)^p{{n-1} \choose p}B_mC_{2n-2p-m}a^m
b^{2n-2p-m}+a-b=1.
\ee
Finally if we denote the internal energy per unit link by
$U_{d,V}$, we have by definition
\be\l{19}
{{d(d-1)}\o 2}U_{d,V}={1\o {2N^2}}
{\partial \o {\partial \b}}{\rm ln} Z_{d,V}(\b,N).
\ee
After some calculation, it can be shown that at large--N limit,
the internal energies in weak and strong regimes for $ V = {\rm
Tr}(AA^{\d})^n $ are:
\be\l{20}
d(d-1)U^{(w)}_{d,n}={{2n-1}\o {\b^{2n}}}\int_a^bz^{2n}\ro_w(z)dz-d
+({1\o n}-2){1\o \b},
\ee
and
\be\l{21}
d(d-1)U^{(s)}_{d,n}={{2n-1}\o {\b^{2n}}}\int_a^b(z^2-a^2)^n\ro_s(z)dz-d
+({1\o n}-2){1\o \b}.
\ee
These equations can be used to deduce the order of phase transition of
the models.

\section{ $d=0$  gSCM with $V = {\rm Tr}(AA^{\d})^n$}

It is clear from the definition of $Z_{d,V}$ in eq.(\r{6}) that
$Z_{0,V}=1$, however ${\tilde Z}_{0,V}$ is not trivial. To see
this, let us focus on $V={\rm Tr}(AA^{\d})^n$ from now on. It is
not difficult to show that ${\tilde Z}_{0,n}$ from eq.(\r{7}) is
\c{Ali}
\be\l{22}
{\tilde Z}_{0,n}\sim {\rm exp}[{N^2(2n-1)\o 2}{\rm ln}\b].
\ee
Now to check the procedure introduced in the last section for
studying the theory in the large--N limit, it is instructive to
reproduce this result by using the density function $\ro(z)$ at
$d=0$.

At $d=0 $, the saddle--point equation (\r{9}) in weak regime
($c=0$), for the case $a_k=\delta_{k,n}$, becomes
\be\l{23}
\frac{n}{2\b^{2n-1}}z^{2n-2}
 = {\cal P}\int_a^b \frac{\ro_w(z')}{z^2-z'^2}dz'\;\;\;\;\;\ ,
 \;\;\;\;\;\; \b>\b_c
\ee
and in the strong regime ($c=a^2$) becomes
\be\l{24}
\frac{n}{2\b^{2n-1}}(z^2-a^2)^{n-1}
 = {\cal P}\int_a^b
 \frac{\ro_s(z')}{z^2-z'^2}dz'\;\;\;\;\ , \;\;\;\; \b<\b_c.
\ee
Let us focus on $\b<\b_c$ case. To make the integral equation (\r{24})
more conventional, we use the change of variable $z \rightarrow \la = z^2$,
with density function $\ro_s(\la)$ satisfies
\be\l{25}
\ro_s(\la)d\la = \ro_s(z')dz'.
\ee
So
\be\l{26}
\frac{n}{2\b^{2n-1}}(\la-a^2)^{n-1}
 = {\cal P}\int_{a^2}^{b^2}
 \frac{\ro_s(\xi)}{\la-\xi}d\xi\;\;\;\;\ , \;\;\;\; \b<\b_c.
\ee
Now consider the function $H_s(\la)$ in complex--$\la$ plane
\be\l{27}
H_s(\la) = \int_{a^2}^{b^2}\frac{\ro_s(\xi)}{\la- \xi}d\xi.
\ee
This function is analytic on the entire complex plane except for a cut
on the positive real axis in the interval $[a^2, b^2]$. Then one has
\be\l{28}
H_s(\la \pm i\epsilon)=R_s(\la) \mp i\pi \ro_s(\la) \;\;\; , \;\;\;
 a^2 \leq \la \leq b^2,
\ee
where $R_s(\la)$ is, from eq.(\r{26}),
\be\l{29}
R_s(\la) = {n \o {2\b^{2n-1}}}(\la - a^2)^{n-1}.
\ee
Using the standard method of solving the integral equations \c{ga}, one can
show that the expression
\be\l{30}
H_s(\la) = {1\o{2\pi i}} \sqrt{\frac{\la-b^2}{\la-a^2}} \oint_c
\sqrt{\frac{\xi -a^2}{\xi-b^2}}\frac{R_s(\xi)}{\la-\xi}d\xi ,
\ee
has the correct analytical behavior in $\b < \b_c$ region (see
\c{Ali} for more details). Here the contour $c$ is a contour
encircling the cut $[a^2, b^2]$ and excluding $\la$. Deforming $c$ to a
contour around the point $\la $ and the contour $c_{\infty}$ (a
contour at infinity), one finds
\be\l{31}
H_s(\la ) = R_s(\la ) + {1\o{2\pi i}}\sqrt{\frac{\la-b^2}{\la-a^2}}
\oint_{c_{\infty}}\sqrt{\frac{\xi-a^2}{\xi-b^2}}\frac{R_s(\xi)}{\la - \xi}d\xi .
\ee
Inserting (\r{29}) in (\r{31}), it can be easily shown that $\ro_s(\la)$
is (using (\r{28})):
\be\l{32}
\ro_s^{(n)}(\la)=-{n\o {2\pi\b^{2n-1}}}\sqrt{{{b^2-\la}\o {\la-a^2}}}\sum_{m,p,s=0}
(-1)^s {{n-1} \choose s}B_pC_{n-p-m-s-1}\la^ma^{2(p+s)}
b^{2(n-p-m-s-1)}.
\ee
At $n=1$, where gSCM reduces to SCM, it can be shown that (\r{32}) is equal
\be\l{33}
\ro^{n=1}_s(\la) = {1\o{2\pi \b}} \sqrt{\frac{b^2-\la}{\la-a^2}},
\ee
which is the same as one calculated in \c{rc3}. To find the parameters $a$ and $b$
in eq.(\r{32}), we note that at $|\la| \rightarrow \infty $, eqs.(\r{27})
and (\r{10}) imply $H(\la ) \rightarrow 1/\la$ or
$\sqrt{(\la-a^2)/(\la-b^2)}H(\la ) \rightarrow 1/\la $.
Therefore we can expand $\sqrt{(\la-a^2)/(\la-b^2)}(R_s(\la )-i\pi \ro_s(\la)) $
and take the coefficients of $\la^0$ and $1/\la$ equal to 0 and 1, respectively.
It can be see that the coefficient of $\la^0$ is identically zero, and the second
condition yields to (for $\b < \b_c$)
\be\l{34}
{n\o{2\b^{2n-1}}}\sum_{p,s=0} (-1)^s{{n-1}\choose s}B_pC_{n-s-p}a^{2(p+s)}
b^{2(n-p-s)} = 1.
\ee
In $n=1$ case, eq.(\r{34}) reduces to $b^2-a^2 = 4\b$, which is one obtained
in \c{rc3}.
The same calculation for the density function $\ro_w(\la)$ yields to
\bea \l{35}
\ro^{(n)}_w(\la) = \left\{ \begin{array}{ll}
0   \;\;\;\;\;\;\;\;\;\;\;\;\;\;\; \hspace{9.0cm} n=1  \\
{n\o{2\pi\b^{2n-1}}}\sqrt{(b^2-\la)(\la-a^2)}\sum_{p,s=0}
 C_pC_{n-p-s-2}\la^sa^{2p}b^{2(n-p-s-2)} \;\;\;\;\;            n>1
\end{array}\right.
\eea
and the following equations which specify the parameters $a$ and
$b$ in $\b>\b_c$,
\be\l{36}
\frac{n}{2\beta^{2n-1}}\sum_{p=0}C_pC_{n-p-1} a^{2p}b^{2(n-p-1)} = 0,
\ee
\be\l{37}
\frac{n}{2\beta^{2n-1}}\sum_{p=0}C_pC_{n-p} a^{2p}b^{2(n-p)} = 1.
\ee
Now let us focus on critical point $\b=\b_c$ in which $a=a_c$ and
$b=b_c$ must satisfy eqs.(\r{34}), (\r{36}) and (\r{37})
simultanously. It is not difficult to see at $\b = \b_c$, except
the first term of eqs. (\r{34}) and (\r{37}) which is equal, all
the other terms are not the same. The only unique solution of
this inconsistency is
\be\l{38}
a_c = 0.
\ee
Inserting (\r{38}) in eqs.(\r{36}) and (\r{37})(or (\r{34})), results
\be\l{39}
\frac{b_c^{2n-2}}{\b_c^{2n-1}}=0,
\ee
\be\l{40}
{n\o{2\b_c^{2n-1}}}C_nb_c^{2n} = 1.
\ee
The solution of these equations are
\bea\l{41}
b_c & \longrightarrow & \infty  ,\nonumber \\
\b_c & \longrightarrow & \infty ,
\eea
such that eq.(\r{40}) holds. Now in the weak--coupling region, where
$\ro_w \sim 1/\b^{2n-1}$, $\b$ is always greater than $\b_c= \infty$, so
$\ro_w \rightarrow 0$. Therefore there is only {\it one} regime at $d=0$, i.e. the
strong--coupling regime.

To check eq.(\r{22}), it is easier first to calculate the contribution of
${\tilde Z}_{0,n}$ to internal energy $U^{(s)}_{0,n}$ and then find
${\tilde Z}_{0,n}$ from it. If we denote this contribution by
${\tilde U}^{(s)}_{0,n}$, it is equal to (see eq.(\r{21}))
\bea\l{42}
{\tilde U}^{(s)}_{0,n} & = & \frac{2n-1}{2\b^{2n}}\int_a^b(z^2-a^2)^n
\ro_s^{(n)}(z)dz \nonumber \\
 & = & \frac{2n-1}{2\b^{2n}}\int_{a^2}^{b^2}(\la-a^2)^n
\ro_s^{(n)}(\la)d\la ,
\eea
in which $\ro^{(n)}_s(\la)$ is given by eq.(\r{32}) and the relation between
$a$ and $b$ can be obtained from (\r{34}). To see the consistency of the
results, it is sufficient to check some specific examples. In $n=2$ ,
(\r{42}) reduces to
\be\l{43}
{\tilde U}^{(s)}_{0,2} = {27\o{256\b^7}}(b^2-a^2)^4,
\ee
and (\r{34}) reduces to
\be\l{44}
{3\o 8}(b^2-a^2)^2 = \b^3 ,
\ee
therefore
\be\l{45}
{\tilde U}^{(s)}_{0,2} ={ 3\o{4\b}} = {1\o {2N^2}} \frac{\partial}{\partial\b}
{\rm ln}{\tilde Z}_{0,2}.
\ee
This equation gives ${\tilde Z}_{0,2} \sim {\rm exp}(\frac{3N^2}{2}{\rm ln}\b)$,
which is the same behavior as eq.(\r{22}). It can be also shown that for
$n=3,4$, and $5$, these two methods have the same result. As another reason
for the fact that only the strong--coupling regime exists in $d=0$, it can
be seen that the ${\tilde Z}_{0,n}$ calculated from $\ro_w(\la)$
(eq.(\r{35})) does not coincide with eq.(\r{22}).

\section{Phase structure of gSCM at $d=2$}

As mentioned earlier, the gSCM can be solved analytically only in
$d=0$ and $d=2$ cases, which the latter is an important case
because of its equivalence to gYM$_2$ (and in special YM$_2$). So in
this section we want to study the gSCM at $d=2$ for $V = {\rm
Tr}(AA^{\d})^n$.

It is clear from eqs.(\r{13}) and (\r{14}) that the density
function are known in the weak and strong regimes at $d=2$, as
the second term in the right--hand sides of these equations
vanishes. It can also be shown that at $d=2$, $\ro_w(0)$ must be
equal to zero at critical point $\b=\b_c$. In this way, one can
obtain the precise value of $a$, $b$, and $\b$ at critical point
as following \c{Ali}:
\bea \l{46}
a_c & = & 0 , \nonumber \\
b_c & = & {2n\o{2n-1}}, \nonumber \\
\b_c & = & {2n\o{2n-1}}{{\Biggr [}\frac{n(4n-3)!!}{2^{2n}(2n-1)!}{\Biggl
]}}^{1\o{2n-1}}.
\eea
To study the phase structure of this model, it is necessary to
calculate the internal energies in both weak and strong regimes.
We know the functional form of the density function in these two
regimes, but there are two unknown parameters $a$ and $ b$ in
each of these densities which must be obtained from the
eqs.(\r{15}) and (\r{16}) for weak regime and from eqs.(\r{17})
and (\r{18}) for strong regime. Unfortunately these equations are
too complicated to be solved exactly, but the crucial point is that as
we want to study the phase structure of the model, it is
sufficient to look at the solutions near the critical point.
Therefore we expand the equations (\r{15}) and (\r{16}) around $a_c=0$
and $b_c={2n/{(2n-1)}}$, and find $a_w$ and $b_w$ in terms of $\a = \b - \b_c$
up to second order. After a lengthy calculation one finds
\bea\l{47}
a_w & = & \frac{4n-3}{2(2n-1)\beta_c} \alpha
-\frac{4n-3}{2n(4n-5)\beta_c^2} \alpha^2 + \ldots \nonumber \\
b_w & = & \frac{2n}{2n-1}+ \frac{4n-1}{2(2n-1)\beta_c} \alpha
-\frac{1}{8n\beta_c^2} \alpha^2 + \ldots
\eea
Now inserting $\ro_w(z)$ from eq.(\r{13}), for $a_m = \delta_{n,m}$
and $d=2$, into eq.(\r{20}), we arrive at
\be\l{48}
 U^{(w)}_{2,n}= -\frac{n(2n-1)}{2\pi\b^{4n-1}} \sum_{p,q,r=0}
 B_rC_pC_{2n-p-q-2} a^{p+r}b^{4n-p-q-2} I(2n+q-r) -1 + ({1\o {2n}}-1){1\o\b},
\ee
in which
\be\l{49}
I(k) = \frac{\sqrt{\pi}b^{2+k}\Gamma(3/2 +k)}{2\Gamma(3+k)}
- \frac{a^{\frac{3}{2}+k}\sqrt{b}}{\frac{3}{2}+ k}\ \ {_2F_1[-\frac{1}{2}\;
, \; {3\o2}+k , \; \frac{5}{2}+k, \; \frac{a}{b}]},
\ee
where ${_2F_1[a,b,c,x]}$ is the hypergeometric function. Keeping
the terms up to second order of $a_w^2$, eq.(\r{48}) becomes:
\be\l{50}
 U^{(w)}_{2,n} = K( T_1b_w^{4n}-{1\over 2} T_2b_w^{4n-1}a_w +
T_{3} b_w^{4n-2}a_w^2 + \ldots) -1 +({1\o{2n}} - 1){1\o{\b}},
\ee
where
$$ K=\frac{n(2n-1)}{4\sqrt{\pi}\beta^{4n-1}}, $$
$$ T_1=\sum_{q=0}C_{2n-q-2}M(n,q,0), $$
$$ T_2=C_{2n-2}M(n,0,1), $$
\be\l{51}
T_{3}=\sum_{q=0}{\Biggr [}{3 \over 8} C_{2n-q-4}M(n,q,0)
-{1 \over 4}C_{2n-q-3}M(n,q,1)-{1 \over 8}C_{2n-q-2}M(n,q,2)
{\Biggl ]} ,
\ee
in which
\be\l{52}
M(n,q,r) = \frac{\Gamma{(3/2 +2n +q -r)}}{\Gamma{(3+2n+q-r)}}.
\ee
In the strong regime, the same calculation leads to
\bea\l{53}
a_s & = & -\frac{4n-3}{2(2n-1)\beta_c}\alpha - \frac{(n-1)(4n-3)}{2n(4n-5)
\beta_c^2}\alpha^2 +\cdots ,\nonumber \\
b_s & = & \frac{2n}{2n-1} + \frac{4n-1}{2(2n-1)\beta_c}\alpha + \frac{n-1}{2n\beta_c^2}
\alpha^2+\cdots ,
\eea
for the parameters $a$ and $b$, and
\be\l{54}
U^{(s)}_{2,n} = K( T_1b_s^{4n}+{1\over 2} T_2b_s^{4n-1}a_s +
S_3 b_s^{4n-2}a_s^2 + \ldots) -1 +({1\o{2n}}-1)\frac{1}{\b},
\ee
for internal energy. $S_3$ in eq.(\r{54}) is
\be\l{55}
S_3 =\sum_{q=0}{\Biggr [}({7 \over 8}-n)C_{2n-q-4}M(n,q,0)
-{1 \over 4}C_{2n-q-3}M(n,q,1)+({3 \over 8}-n)C_{2n-q-2}M(n,q,2){\Biggl ]}.
\ee
Now subtracting the internal energies in two regions (eqs.(\r{54}) and
(\r{50})) and using eqs.(\r{47}) and (\r{53}), one obtains
\be\l{56}
U^{(s)}_{2,n} - U^{(w)}_{2,n} = f(n)(\b - \b_c)^2 + \ldots  ,
\ee
where
\be\l{57}
f(n) = \frac{(2n-1)(4n-3)b_c^{4n-1}}{32\sqrt{\pi}\beta_c^{4n+1}}
{\Biggr \{} 4nT_1 - \frac{4n-3}{2(4n-5)}T_2 - \frac{4n-3}{2}T_4 {\Biggl \}},
\ee
in which
\be\l{58}
T_4 = C_{2n-2}M(n,0,2) + C_{2n-3}M(n,0,1) +2 \sum_{q=0}C_{2n-q-4}M(n,q,0).
\ee
Eq.(\r{56}) shows that all the $V={\rm Tr}(AA^{\d})^n$ gSCM has a
{\it third order} phase transition near the critical point $\b = \b_c$.
Note that $f(n) >0$ for all values of $n$, as we expected from the
definitions of strong and weak regimes, and also the value  of $f(n)$
for $n = 2,3,$ and $4$ coincides with those in ref.\c{Ali}.

As a
final comment, it is interesting to compare the behavior of the
solutions with respect to $n$. The only physical comparable
quantity, in the case of third order phase transition, is the
difference
\be\l{59}
\frac{\partial C_s}{\partial T}-\frac{\partial C_w}{\partial T}=
\b^2\frac{\partial}{\partial \b}{\Biggr (}\b^2\frac{\partial}{\partial \b}
(U^{(s)}-U^{(w)}){\Biggl )},
\ee
where $C$ is the specific heat of the theory and $T = {1/\b}$ is
temperature. Using (\r{56}), it is found
\be \l{60}
\b^2\frac{\partial}{\partial \b}(\Delta C)_n = 2\b_c^4 f(n).
\ee
In Fig. (1), where  we have plotted $\b^2\frac{\partial}{\partial \b}(\Delta C)_n$
vs $n$, one can see a saturating value 2 for this difference.


\begin{thebibliography}{99}
\bibitem{Ali} M. Alimohammadi; Nucl. Phys. {\bf B565} (2000)
469.
\bibitem{gt} G. 't Hooft; Nucl. Phys. {\bf B72} (1974) 461.
\bibitem{dg} D. J. Gross and E. Witten; Phys. Rev. {\bf D21} (1980) 446.
\bibitem{rc1} R. C. Brower and M. Nauenberg; Nucl. Phys. {\bf B180} (1981) 221.
\bibitem{eb} E. Brezin and D. J. Gross; Phys. Lett. {\bf B97} (1980) 120.
\bibitem{rc2} R. C. Brower, P. Rossi and C. -I. Tan; Phys. Rev. {\bf D23}
(1981) 942; {\bf D23} (1981) 953.
\bibitem{df} D. Friedan; Commun. Math. Phys. {\bf 78} (1981) 353.
\bibitem{Cam} M. Campostrini, P. Rossi, and E. Vicari, Phys. Rev. {\bf D52}
(1995) 358.
\bibitem{pr2} P. Rossi and C. -I. Tan; Phys. Rev. {\bf D51} (1995) 7159.
\bibitem{rc3} R. C. Brower, M. Campostrini, K. Orginos, P. Rossi, C. -I. Tan,
and E. Vicari; Phys. Rev. {\bf D53} (1996) 3230.
\bibitem{ew} E. Witten; Jour. Geom. Phys. {\bf 9} (1992) 303.
\bibitem{mr} M. R. Douglas, K. Li and M. Staudacher; Nucl. Phys. {\bf B240}
             (1994) 140;\\
             O. Ganor, J. Sonnenschein, and S. Yankielowicz;
             Nucl. Phys. {\bf B434} (1995) 139.
\bibitem{mk} M. Khorrami and M. Alimohammadi; Mod. Phys. Lett. {\bf A12} (1997)
             2265.
\bibitem{ma1} M. Alimohammadi, M. Khorrami and, A. Aghamohammadi;
             Nucl. Phys. {\bf B510} (1998) 313.
\bibitem{ma2} M. Alimohammadi and A. Tofighi;
 Eur. Phys. Jour. {\bf C8} (1999) 711.
\bibitem{ga} D. Gakhov, ``Boundary problems" , Russian edition
(Nauka, 1975), A. C. Pipkin; ``A course on integral equations"
(Springer, Berlin, 1991).
\end{thebibliography}
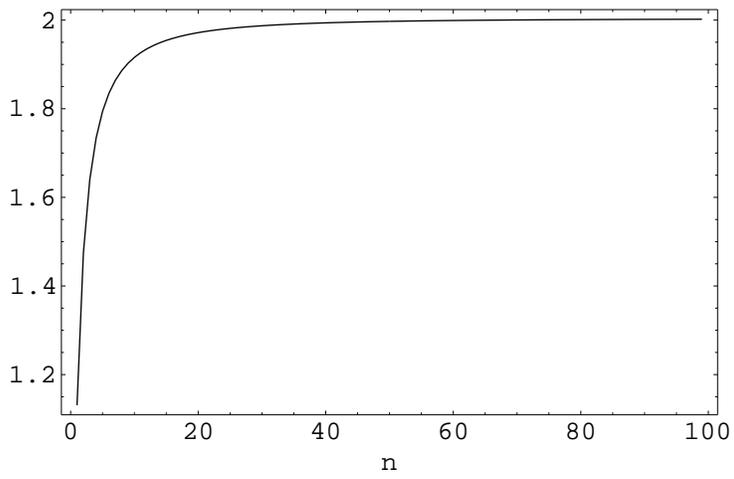
\end{document}